\documentstyle[12pt]{article}
\newcommand{\beq}{\begin{equation}}
\newcommand{\eeq}{\end{equation}}
\newcommand{\bea}{\begin{eqnarray}}
\newcommand{\eea}{\end{eqnarray}}
\begin{document}

\noindent
{\Large{\bf Multi-connected momentum distribution and fermion condensation}}

\vskip 0.5 cm

\noindent
\hskip 2.7 cm 
M V Zverev$\dag$ and M Baldo$\ddag$

\begin{small}

\vskip 0.4 cm
\noindent
\hskip 2.7 cm $\dag$ Kurchatov Institute, 123182 Moscow, Russia

\noindent
\hskip 2.7 cm $\ddag$ Istituto Nazionale di Fisica Nucleare,
Corso Italia 57, 95129 Catania, Italy

\vskip 0.6 cm

\hangindent=2.7 true cm \hangafter=0 \noindent
{\bf Abstract.} The structure of the ground state beyond the instability point
of the quasiparticle system with Fermi-step momentum distribution
is studied within the model of a Fermi liquid with a strong repulsive 
interaction. A ground state rearrangement occurs as the interaction 
strength is increased beyond a definite critical value.
Numerical investigation of the initial stage of this structural transition
shows that there are two temperature regions, corresponding to different 
scenarios of the rearrangement. 
While for temperature $T$ larger than some characteristic
temperature $T_0$ the
behaviour of the system is the same as that in the case of the 
fermion condensation, for $T{<}T_0$ the intermediate
structure with multiconnected quasiparticle momentum distribution
arises. The transition of this structure to the fermion condensate 
at increasing interaction strength is discussed.  



\end{small}

\vskip 0.7 cm

\noindent
{\bf 1.~Introduction}
\vskip 0.3 cm
 
\noindent

The question of an applicability of the Landau Fermi liquid
theory \cite{landau} for describing properties of strongly correlated
Fermi systems has been discussed since a long time. It is known
that this theory is not valid for one dimensional (1D) systems 
\cite{luttinger}. For such systems,
the concept of Luttinger liquid \cite{haldane} with a single-particle
Green function containing no quasiparticle pole is usually introduced, 
instead of the quasiparticle picture. The frontiers of the non-Fermi liquid
view cover also strongly correlated 2D liquids 
\cite{anderson,varma}, since HTC
materials with quasi-2D structure possesse properties in 
contradiction with the expected ones in
the Landau theory. However the recently measured electronic spectra of
such materials \cite{shen,yokoya,lu} show evidence for the presence of 
a quasiparticle pole
in the single-electron propagator.
At the same time, new possibilities were found within the
quasiparticle approach in refs.~[9--12], 
where quasiparticles
with momentum distributions differing from the ones assumed in the Landau 
theory are introduced. This new class of systems with the presence of a 
fermion condensate, as predicted in refs.~\cite{ks,ksk}, possesses a rich
variety of properties [12--15], some of which are
characteristics of a non-Fermi liquid behaviour. As discussed within
different models [12--14,~16], a state with the fermion
condensate arises as a result of the rearrangement of the ground
state of the system. This rearrangement in a system of
quasiparticles, whose momentum distribution has the shape of a
Fermi sphere with an occupation number slightly smoothed at $T{>}0$,
takes place when some parameters are varied and
the pertinent stability condition is violated. In the present
work we consider the model of a homogeneous Fermi liquid, which
also displays a rearrangement of the ground state of the
quasiparticle system, and investigate the scenario of the initial
stage of the rearrangement.

\vskip 0.5 cm
\noindent
{\bf 2.~Theoretical and computational aspects}

\vskip 0.3 cm
\noindent
{\it 2.1.~Two-connected Fermi sphere and the fermion condensate}

\vskip 0.3 cm
\noindent
Let us start by recalling the Landau relation between
the quasiparticle momentum distribution $n_{\bf p}(T)$ and
the quasiparticle spectrum $\varepsilon_{\bf p}(T)$,
\beq  
n_{\bf p}(T)=
\left\{1{+}\exp{\varepsilon_{\bf p}(T){-}\mu(T)\over T}
                                 \right\}^{-1},
\eeq
($\mu(T)$ is the chemical potential), which results from
a variational equation
$\delta F/\delta n_{\bf p}=0$
($F$ is the free energy of the system) with th usual
expression for the entropy \cite{three}. On one hand, eq.~(1) is
simply the Fermi-Dirac quasiparticle distribution over energies.
On the other hand, this relationship is an equation for the quasiparticle
distribution over the phase space. In fact, the quasiparticle energy is
the variational derivative of the ground state energy
functional $E_0[n]$ with respect to the quasiparticle distribution,
$\varepsilon_{\bf p}(T)=\delta E_0/\delta n_{\bf p}(T)$, and, therefore,
it is itself a functional of $n_{\bf p}(T)$.

It is postulated in the Landau theory that in a homogeneous isotropic
Fermi liquid, like in a Fermi gas, the quasiparticle momentum
distribution at $T{=}0$ has the shape of a fully occupied Fermi sphere
$n_F^{(0)}(p){=}\theta(p_F{-}p)$ (the maximum momentum $p_F$ is
related to the density $\rho$ of the system by the well known relation
$\rho{=}p_F^3/(3\pi^2)$). The low-temperature behaviour of the
quasiparticle spectrum corresponding to such a momentum distribution
has the form \cite{three}:
\beq  
\varepsilon_p(T)-\mu(T)=\xi(p)+O(T^2).
\eeq
The function $\xi(p)$ increases monotonically in the vicinity of the
Fermi momentum and changes its sign at $p{=}p_F$. Its slope at this
point, which is the group velocity of the quasiparticles on the Fermi surface
$v_F{=}d\xi(p)/dp|_{p{=}p_F}$, is determined by one of the phenomenological
parameters of the Fermi liquid theory, the effective mass
$M^*{=}p_F/v_F$.

In a strongly correlated Fermi system, a quasiparticle momentum
distribution minimizing the energy functional $E_0[n(p)]$ at $T{=}0$
may be located out of the corner point $n_F^{(0)}(p)$ of the
functional space $[n]$ and the low-temperature behaviour of the
corresponding quasiparticle spectrum may differ from eq.~(2).
For instance, it was found in ref.~\cite{gilbert} that quasiparticle 
energies which are
equal to the chemical potential in a finite region of the momentum
space can exist. In refs.~\cite{llano1,llano2}, some model functionals 
$E_0[n(p)]$ were introduced which, for certain values of the parameters, 
reach their absolute minimum for a momentum distribution
characterized by a two-connected Fermi sphere
\beq 
n_F^{(1)}(p)=
   \theta(p_1{-}p)-\theta(p_2{-}p)+\theta(p_3{-}p).
\eeq

A quite different quasiparticle ground state corresponds to systems
with fer\-mi\-on condensate [11--14]. 
Let us elucidate the main idea
of the concept of a fermion condensate. A homogeneous and
isotropic quasiparticle system with the fermion condensate is
described by a singular solution of eq.~(1) which corresponds
to a quasiparticle spectrum linear in the temperature $T$ within
a finite region of momenta \cite{ks,ksk}:
\beq  
\varepsilon_p(T)-\mu(T)=T\nu_0(p)+o(T), \qquad p_i{<}p{<}p_f.
\eeq

Contrary to the Fermi liquid formula (2), there is no 
$T$-independent term in eq.~(4). This means that at $T{=}0$ 
the quasiparticle spectrum has a plateau $\varepsilon_p\equiv\mu$ 
in the region $p_i{<}p{<}p_f$. At $T{>}0$ the slope of the plateau 
is linear in $T$, and its position with respect to the chemical
potential $\mu(T)$ is determined by the function $\nu_0(p)$
which is connected with the momentum distribution of quasiparticles
in the condesate. Indeed, the singular solution of eq.~(1), which
can be easily obtained upon substitution the formula (4) into eq.~(1),
has the form $n_p(T){=}n_0(p){+}O(T)$, where
\beq  
n_0(p)=\Bigl\{1{+}\exp(\nu_0(p))\Bigr\}^{-1}, \qquad p_i{<}p{<}p_f
\eeq
is the momentum distribution of the condensate quasiparticles at
$T{=}0$. Outside the condensate region, $n_0(p){=}1$ at $p{<}p_i$,
and $n_0(p){=}0$ at $p{>}p_f$ \cite{ks,ksk}. One can find the explicit
form of $n_p(T)$ and $\varepsilon_p(T)$ provided one knows the
functional dependence $\varepsilon_p(T)[n_p(T)]$. A set of model
functionals was suggested in refs.~[12--14,~16], each of them
possessing the minimum at the singular solution, within well defined
regions of functional parameters. In the 
present work we indicate the possibility of a scenario, in
which the transition to the fermion
condensate occurs through an intermediate structure, corresponding to
a multiconnected quasiparticle momentum distribution.

\vskip 0.5 cm
\noindent
{\it 2.2.~The effective functional}

\vskip 0.3 cm
\noindent
The initial stage of the rearrangement, which is studied in this 
paper, is characterized by variations of the quasiparticle
momentum distribution within a relatively thin layer around
$p{\sim}p_F$ (see results below). Under these conditions one can
use the concept of effective functional, which is widely 
used in many-body theory. We consider the simple effective 
quasiparticle functional for the ground state energy:
\beq  
E_0[n_p(T)]=\int\frac{p^2}{2M}\,n_p(T)\,d\tau +
\frac{1}{2}\int V(\bf p{-}\bf p')\,n_p(T)\,n_{p'}(T)\,
d\tau\,d\tau'
\eeq
with the effective repulsive interaction
\beq  
V({\bf p}{-}{\bf p'}) = \frac{V_0}{({\bf p}{-}{\bf p'})^2+\alpha^2}.
\eeq
The symbol $d\tau$ in eq.~(6) means integration over
$d^3p'/(2\pi)^3$ and summation over spin indices.
Calculating the variatonal derivative of $E_0$ with respect
to quasiparticle distribution $n_p(T)$ one obtains the
quasiparticle spectrum
\beq  
\varepsilon_p(T)=\frac{p^2}{2M} +
\int V({\bf p}{-}{\bf p'})\,n_{p'}(T)\,d\tau.
\eeq
The functional dependence given by eqs.(7),~(8) together
with eq.~(1) and the normalization condition
\beq  
\int n_p(T) d\tau = \rho
\eeq
are the set of equations to be solved for the quasiparticle
disrtibution $n_p(T)$ and the spectrum $\varepsilon_p(T)$.
The value $\alpha{=}0.07p_F$ was used in the calculations of
the present work and the behaviour of the system as the
parameter $V_0$ is varied was studied. The dimensionless
parameter $\gamma{=}MV_0/(4\pi^2p_F)$ will be used in the
discussion below.

\vskip 0.5 cm
\noindent
{\it 2.2.~Numerical aspects}

\vskip 0.3 cm
\noindent
Eq.~(8) together with the expression (1) represents the
nonlinear integral equation for the function $\varepsilon_p(T)$.
This equation was solved numerically by an iterative
procedure with weighting factors. The 5-point
Newton-Cotes quadrature formula with 5-point filter on output
was used for numerical folding the distribution $n_p(T)$ with
the effective interaction $V(p,p')$. The momentum grid had a step
$h_p{=}5{\cdot}10^{-5}p_F$. The accuracy of the numerical solution
was determined upon substitution it into the initial equation.
The permissible error, that is the maximum discrepancy between
the left and the right hand sides of eq.~(8), was fixed at
$10^{-8}\varepsilon_F$. The number of iterations necessary
for reaching this accuracy is about $3\cdot 10^4$ provided
the iteration weight $w{=}0.001$ is taken, which is optimal
for a stability of the iterative procedure. It is worth to note
that the results are independent of the point in the functional
space taken as an initial one for the the iterative procedure. For example,
the same solution for $\gamma{=}0.50$ at $T{=}10^{-7}$ was
obtained by starting from i) the solution for $\gamma{=}0.50$
at $T{=}10^{-5}$, ii) the solution for $\gamma{=}0.48$
at $T{=}10^{-7}$ (here and below the temperature $T$
is taken in units of $\varepsilon_F^0=p_F^2/2M$).

\vskip 0.5 cm
\noindent
{\bf 3.~Results}

\vskip 0.3 cm
\noindent
{\it 3.1.~Instability of the Fermi-step momentum distribution}

\vskip 0.3 cm
\noindent
We begin from estimating the value $\gamma_c^{(0)}$ 
of the interaction strength at which the
necessary condition for stability of the system with the quasiparticle
distribution $n_F^{(0)}(p)$ at $T{=}0$ is violated.
 This condition is
fulfilled \cite{ks,ksk} provided the variation of the ground state energy $E_0$
under any admissible variations of the distribution $n(p)$,
\beq  
\delta E_0 = \int \Bigl[\varepsilon(p){-}\mu\Bigr]\,
\delta n(p)\, \frac{d^3p}{(2\pi)^3},
\eeq
is positive. Admissible variations of $n_F^{(0)}(p)$ are of the same
sign as the difference $p{-}p_F$, as dictated by Pauli principle. 
Hence, upon substitution of the energy
$\varepsilon(p_F)$ for the chemical potential $\mu$ in eq.~(10)
one can reformulate the necessary stability condition as a
requirement for the value
\beq  
s(p) = 2M\, \frac{\varepsilon(p){-}\varepsilon(p_F)}
                 {p^2-p_F^2}
\eeq
to be positive for each value of the momentum $p$ \cite{ks,ksk}.
Therefore, if the function $s(p)$ has the first zero close to $p_F$,
the violation of the stability condition means that
a bend appears in the curve $\varepsilon(p)$ in the vicinity of
the Fermi momentum. For the distribution $n_F^{(0)}(p)$ the derivative
$d\varepsilon/dp$ can be easily calculated in the dimensionless form
\beq  
\zeta^{(0)}(p) = \frac{M}{p_F}\frac{d\varepsilon}{dp} =
        \frac{p}{p_F} + \frac{\gamma p_F}{p}
 - \frac{\gamma (p^2{+}p_F^2{+}\alpha^2)}{4p^2}\,
  \ln\frac{(p{+}p_F)^2{+}\alpha^2}{(p{-}p_F)^2{+}\alpha^2}.
\eeq

In fig.~1, where the curves $\zeta^{(0)}(p)$ are shown for
different values $\gamma$, one can see that the contact
with zero takes place at the point $p=p_c\simeq 0.97p_F$ for
$\gamma{=}\gamma_c^{(0)}{\simeq}0.415$. Note that the proximity
of $p_c$ to $p_F$ justifies the substitution of the function $s(p)$ 
with the function
$\zeta^{(0)}(p)$. Therefore, the ground state with the quasiparticle
momentum distribution $n_F^{(0)}(p)$ becomes unstable at
$\gamma{>}\gamma_c^{(0)}$ and its rearrangement takes place.

\vskip 0.3 cm
\noindent
{\it 3.2.~Multiconnected momentum distribution}

\vskip 0.3 cm
\noindent
To see how the ground state is arranged right beyond the
transition point, let us look at the fig.~2, where the results
of the calculations of $n_p(T)$ for $\gamma{=}0.45$ at different
$T$ are shown. At $T{=}2{\cdot}10^{-3}$, the quasiparticle momentum
distribution $n_p(T)$ has the shape characteristic of a system with 
fermion condensate (we shall discuss that in detail below). At
$T{<}2{\cdot}10^{-3}$ a downfall appears in the distribution,
which deepens as temperature decreases, and finally, at $T{=}10^{-7}$,
one can hardly distinguish $n_p(T)$ from the two-connected Fermi sphere
defined in eq.~(3). The quasiparticle spectrum $\varepsilon_p(T)$
corresponding to such distribution calculated at $T{=}10^{-7}$
is shown in fig.~3.  The spectrum of a fermion condensate
has a shape with a plateau at a value exactly equal to the chemical
potential $\mu$ at $T{=}0$ and getting little sloping at $T{>}0$
\cite{ks,ksk}. Here we have a quite different behaviour, and
the quasiparticle spectrum of the two-connected Fermi-like
distribution equals $\mu$ at three points, which are the border $p_1$
of the inner sphere and the borders $p_2$ and $p_3$ of the spherical
layer.
The deviation of $\varepsilon_p(T)$ from $\mu$ reaches the value
$\sim 2\cdot 10^{-4}\varepsilon_F$ at the point of the minimum of the
spectrum and the value $\sim 2\cdot 10^{-6}\varepsilon_F$ at the
point of its maximum. Despite the latter value is small, it is still 
higher than the estimated numerical error by two orders of magnitude.

The behaviour of the two-connected Fermi sphere with increasing
interaction strength $\gamma$ is displayed in fig.~4. The spherical
layer appearing beyond the transition point has, from the start, a finite
thickness, while the gap between the layer and the inner sphere develops
starting from a vanishing small width. The outer layer gets thicker and 
moves away 
from the inner sphere as the parameter $\gamma$ increases. What then happens
with such a distribution? To understand that, let us investigate
the stability of such Fermi sphere divided into two layers. We calculate
the function $\zeta^{(1)}(p)$ for the momentum distribution in the form
of eq.~(3) and see when and where the critical change of the sign of that
function occurs. An elementary calculation gives
\beq  
\zeta^{(1)}(p) = \frac{M}{p_F}\frac{d\varepsilon}{dp} =
        \frac{p}{p_F} +
        \sum\limits_{i=1}^3 (-1)^{i-1}\biggl\{
        \frac{\gamma p_i}{p}
 - \frac{\gamma (p^2{+}p_i^2{+}\alpha^2)}{4p^2}\,
  \ln\frac{(p{+}p_i)^2{+}\alpha^2}{(p{-}p_i)^2{+}\alpha^2}
  \biggr\}.
\eeq
The function $\zeta^{(1)}(p)$ calculated for different values of $\gamma$
is displayed in fig.~4. The points of the maximum absolute values of the
derivative $dn/dp$ at $T{=}10^{-7}$ were taken as the boundary momenta
$p_1, p_2, p_3$ for that calculation. One can easily realize that inside
the region $\gamma_c^{(0)}<\gamma<\gamma_c^{(1)} \simeq 0.452$,
the two points where  $\zeta^{(1)}(p)$ changes sign are located
in such a way that the corresponding local minimum and maximum of
$\varepsilon(p)$ lie in the domains where $n(p)$ equals 0 and 1
respectively. This means that the sign of the difference
$\varepsilon(p){-}\mu$
coincides with that of the possible variations $\delta n(p)$ allowed by 
Pauli principle. Hence the considered distribution satisfies the 
stability condition.

However one can see in fig.~4 that at $\gamma >\gamma_c^{(1)} \simeq 0.452$,
the situation changes. With the displayed behaviour of the function
$\zeta^{(1)}(p)$, there are regions where $\varepsilon(p){-}\mu{>}0$,
but with $n(p)=1$. This means that the necessary stability condition is
violated, since allowed variations $\delta n(p)$ exist which lower the
ground state energy. This results in the second rearrangement of the
ground state of the system emerging at $\gamma{=}\gamma_c^{(1)}$.
It is shown in fig.~5 how the quasiparticle distribution is arranged
beyond the second transition point $\gamma_c^{(1)}$. For 
definitness, the results 
of the calculations for $\gamma{=}0.46$ at different $T$ are diplayed. 
The calculations
show that the new layer appears for the above mentioned value of the coupling
constant and the quasiparticle distribution $n_p(T)$ at $T{=}10^{-7}$
is very close to the three-connected Fermi sphere
\beq 
n_F^{(2)}(p)=
   \theta(p_1{-}p)-\theta(p_2{-}p)+\theta(p_3{-}p)
   -\theta(p_4{-}p)+\theta(p_5{-}p).
\eeq
The calculation shows that the appearence of new layers at
increasing values of the parameter $\gamma$ does not stop at
the level of three-connected
Fermi sphere. Figs.~6 and 7, where the quasiparticle distributions
$n_p(T)$ are calculated for $\gamma=0.48$ and $0.50$, show further 
divisions into layers of the momentum space. In particular, 
the distribution $n_p(T)$ for
$\gamma{=}0.50$ approaches a multi-connected Fermi sphere with lowering
$T$, which is arranged as follows: the inner small occupied sphere with
the radius $\sim 0.85 p_F$ is surrounded by four spherical occupied layers
with thickness $\sim 0.3 - 0.4 p_F$ divided by spherical empty layers
with thickness $\sim 0.1 - 0.2 p_F$. The quasiparticle spectra
$\varepsilon_p(T)$ for $\gamma{=}0.50$ are shown in fig.~8.
At $T{=}10^{-7}$,
the spectrum crosses the line $\varepsilon{=}\mu$ nine times, at the
boundary of the inner sphere and at the boundaries of the spherical layers.

\vskip 0.5 cm
\noindent
{\bf 4.~Discussion}

\vskip 0.3 cm
\noindent
{\it 4.1.~Two scenarios of the rearrangement}

\vskip 0.3 cm
\noindent
Thus beyond the first transition point $\gamma_c^{(0)}$, the scenario
of the rearrangement of the quasiparticle ground state at low $T{<}10^{-3}$
is the succession of transitions with increasing $\gamma$. Each one of them
results in the appearance of a new spherical layer in  momentum space.
Let us compare this scenario of the rearrangement and some distinctive
features of the ground state under consideration with those of the
fermion condensation.

Let us recall the main features of the fermion
condensation phenomenon. One of them is the plateau in the quasiparticle 
spectrum
$\varepsilon_p(T)$, with a value exactly equal to the chemical potential 
$\mu$ at
$T{=}0$, in accordance with eq.~(4), and which gets 
a finite slope at increasing
temperature. Unlike systems with the fermion condensate, the spectrum
$\varepsilon_p(T{=}0)$ for the multi-connected Fermi-like distribution
equals the chemical potential in a finite number of points, which
are the boundaries of the spherical layers. The low-temperature
expansion of such a spectrum has the form of eq.~(2), typical of
the Landau theory, with the non-monotone function $\xi(p)$ changing its
sign several times, unlike the monotone one for the usual Fermi liquid.
At $T{=}0$, the ground state with the multi-connected Fermi-like
quasiparticle distribution is not macroscopically
degenerated, unlike the ground state of a fermion condensate.
At the same time, there exist singularities of the density
of states connected with the maxima and the minima of the function
$\varepsilon_p(T{=}0)$. These singularities gradually disappears with
increasing $T$ up to $T_0\sim 2{\cdot}10^{-3}$, where the last
twist of the spectrum corresponding to the outer layer is smoothed.
At $T{>}T_0$ the difference $\varepsilon_p(T)-\mu(T)$ becomes linear
in temperature, like that for systems with a fermion condensate.

The other feature of systems with fermion condensate is the
shape of the distribution $n(p)$ given by eq.~(5).
Within the region occupied by the fermion condensate
$0{<}n(p){<}1$, this corresponds to non-zero entropy of the
fermion condensate at $T{=}0$. The contradiction with the Nernst
theorem disappears providing correlations (e.g. superfluid)
are taken into account, which immediately rearrange the ground state
due to its degeneracy and re-establishes a zero entropy at $T{=}0$.
The entropy of the state with the multi-connected Fermi-like
quasiparticle momentum distribution
equals zero at zero temperature because $n(p)$ takes only the values
0 and 1. The multi-layered distribution changes quickly with
increasing $T$: the sharp boundaries of the layers are smoothed
and the layers combine all together at $T{=}T_0$ into a monotonically
decreasing curve which is very similar to the momentum distribution
of a system with the fermion condensate at that temperature.
Being equal zero at $T{=}0$, the entropy of the system with the
multi-connected Fermi-like distribution increases sharply with temperature
due to quick smoothing of the function $n_p(T)$. The calculation show that
at $T{=}T_0$ the entropy reaches the value
$S_0{\sim}\Omega_0/\Omega$, the ratio of the phase volume of
multi-layered region $\Omega_0$ to that of the whole system
$\Omega$. Just this value of the entropy would be characteristic
at $T{\sim}0$ for a system with the fermion condensate occupying
the phase volume $\Omega_0$. At $T{>}T_0$ the entropy becomes
linear in temperature, like that of a system with a fermion
condensate \cite{ksk,nozieres}.

All these features of the momentum distribution,
entropy, quasiparticle spectrum and density of states for
a system with a multi-connected Fermi-like quasiparticle
distribution, as well as the problem of validity of such a
quasiparticle pattern, will be studied in detail in a separate paper.

The scenario of the fermion condensation is characterized by the
single critical value $\gamma_c$ of the coupling constant,
at which the fermion condensate arises in the system. The phase
volume of the fermion condensate increases as $\gamma$ further
increases, however the shape of the momentum distribution does
not modify and no further qualitative changes occur \cite{ks,ksk}.
The scenario of the rearrangement found for the model under
consideration is different for different temperatures.
At $T{=}0$ it is characterized by a sequence of critical
values $\gamma_c^{(i)}$, each one corresponding to the appearence
of a new ground state with a larger number of connectivity.
The number of critical constants decreases as $T$ increases,
so that only one of them, $\gamma_c^{(0)}$, survives at $T{>}T_0$.
This means that for the considered model, the scenario of formation
of the multi-connected Fermi sphere at $T{=}0$ gradually transforms into
that of the fermion condensation with increasing temperature.

\vskip 0.3 cm
\noindent
{\it 4.2.~Interplay between the multiconnected distribution 
and the fermion condensation}

\vskip 0.3 cm
\noindent
Unfortunately, the computation time sharply increases with increasing
the phase volume $\Omega_0$ of the layered distribution. This is
the reason why 
the calculations of the present work are carried out up to
the value of the interaction strength $\gamma{\le}0.5$, corresponding to 
the initial
stage of the rearrangement. What happens in the system at larger values
of $\gamma$? To have a feeling of that, let us try to use the mechanical 
analogy, treating the momentum $p$ as a spatial coordinate $r$ 
\cite{ks,ksk}. Then the problem of
minimizing the ground state energy functional (6) at $T{=}0$ can be
interpreted in the language of mechanics  as a search for
the statical equilibrium of the spatial distribution $\nu(r)$ of particles,
which move in the external harmonic field $U(r)=kr^2/2$ with the stiffness
$k{=}1/M$ and interact among each others by the repulsive force (7),
the particle number being fixed by the normalization condition (9).
Unless the solution $\nu(r)$ of the mechanical problem exceeds
$2/(2\pi)^3$, at least in one point, it can not be accepted as a solution
$n(p)$ because the latter should satisfy the additional restriction
due to the Pauli principle. With strengthening interparticle repulsion,
the considered mechanical system, obviously, expands and rarefies.
As soon as the distribution $\nu(r)$ is everywhere less than $2/(2\pi)^3$,
one can conclude that it corresponds to the solution $n(p)$
of the initial quantum problem.
Since one can expect that the distribution $\nu(r)$ should be smoothed
and monotone, the fermion condensate seems to appear in the system
at $T{=}0$ for large values of the interaction strength $\gamma$.
This transition will be investigated in further publications.

\vskip 0.5 cm
\noindent
{\bf 5.~Conclusion}

\vskip 0.3 cm
\noindent
In summary, the structure of the ground state of the homogeneous Fermi 
liquid with the strong interparticle repulsion is studied within the 
framework of the effective functional approach. The numerical 
investigation of the simple functional with strong repilsive effective 
interaction showed that at fixed value of the interaction radius there
exist the critical value of the interaction strength $\gamma_c^{(0)}$,
beyond which the ground state with the Fermi-step quasiparticle
momentum distribution becomes unstable and the rearrangement
of the ground state takes place. The scenario of
the initial stage of that rearrangement with increasing $\gamma$ was
found to be different for different regions of temperature.
At $T{=}0$, there exist the set of critical constants
$\gamma_c^{(i)}$ corresponding to the succession of transitions,
each resulting in emerging the new spherical layer of
the quasiparticle momentum distribution $n(p))$.
The quasiparticle spectrum $\varepsilon(p)$ corresponding
to such the layered distribution does not have the
plateau, unlike the fermion condensate, and equals the chemical
potential in the finite number of points, which are the boundaries of
the layers. The ground state with the multi-connected Fermi-like
distribution possesses no macroscopic degeneracy and the entropy of
this state is zero at zero temperature. With increasing temperature
the layers are quickly smoothed, so that at
$T{\sim}T_0{\sim}2{\cdot}10^{-3}$, there is no more reminiscence
of the critical constants $\gamma_c^{(i)}$ with the exception of
$\gamma_c^{(0)}$. At $T{>}T_0$ the scenario of the rearrangement
is that of the fermion condensation. The qualitative analysis 
showed that the found stucture with the multiconnected quasiparticle 
momentum distribution is the intermediate one, yielding the place 
for the fermion condensate with incresing the interaction strength.

\vskip 0.5 cm
\noindent
{\bf Acknowledgments}

\vskip 0.3 cm
We are indebted to V~A~Khodel for the unceasing interest for the 
present research and countless fruitful discussions as well as 
S~A~Artamonov, A~E~Bulatov, E~E~Saperstein, V~R~Shaginyan and 
S~V~Tolokonnikov for valuable discussions. 
This research was partially supported by Grant No.$\,$96-02-19293
from the Russian Foundation for Basic Research. 
M~V~Z would like to thank INFN (Catania, Italy), where the main 
part of the work was done, for the kind hospitality.

\newpage

\vskip 0.5 cm
\noindent
{\bf Figure captions}

\vskip 1 cm
\parindent=0 cm

1. The function $\zeta^{(0)}(p)$ calculated for
$\gamma=0.410$, 0.415, 0.420.

\vskip 0.3 cm
2. The quasiparticle momentum distributions $n_p(T)$ calculated
for $\gamma{=}0.45$ at different temperatures.

\vskip 0.3 cm
3. The quasiparticle spectrum $(\varepsilon_p(T){-}\mu)/\varepsilon_F^0$
calculated for $\gamma{=}0.45$ at $T{=}10^{-7}$.

\vskip 0.3 cm
4. The quasiparticle momentum distributions $n(p)$ and the function
$\zeta^{(1)}(p)$ calculated for different values of the parameter
$\gamma$.

\vskip 0.3 cm
5. The same as for Fig.~2, for $\gamma{=}0.46$.

\vskip 0.3 cm
6. The same as for Fig.~2, for $\gamma{=}0.48$.

\vskip 0.3 cm
7. The same as for Fig.~2, for $\gamma{=}0.50$.

\vskip 0.3 cm
8. The quasiparticle spectra $(\varepsilon_p(T){-}\mu)/\varepsilon_F^0$
calculated for $\gamma{=}0.50$ at $T{=}10^{-4}$ (solid line),
$T{=}10^{-5}$ (long dashes), $T{=}10^{-6}$ (short dashes),
and $T{=}10^{-7}$ (dots).

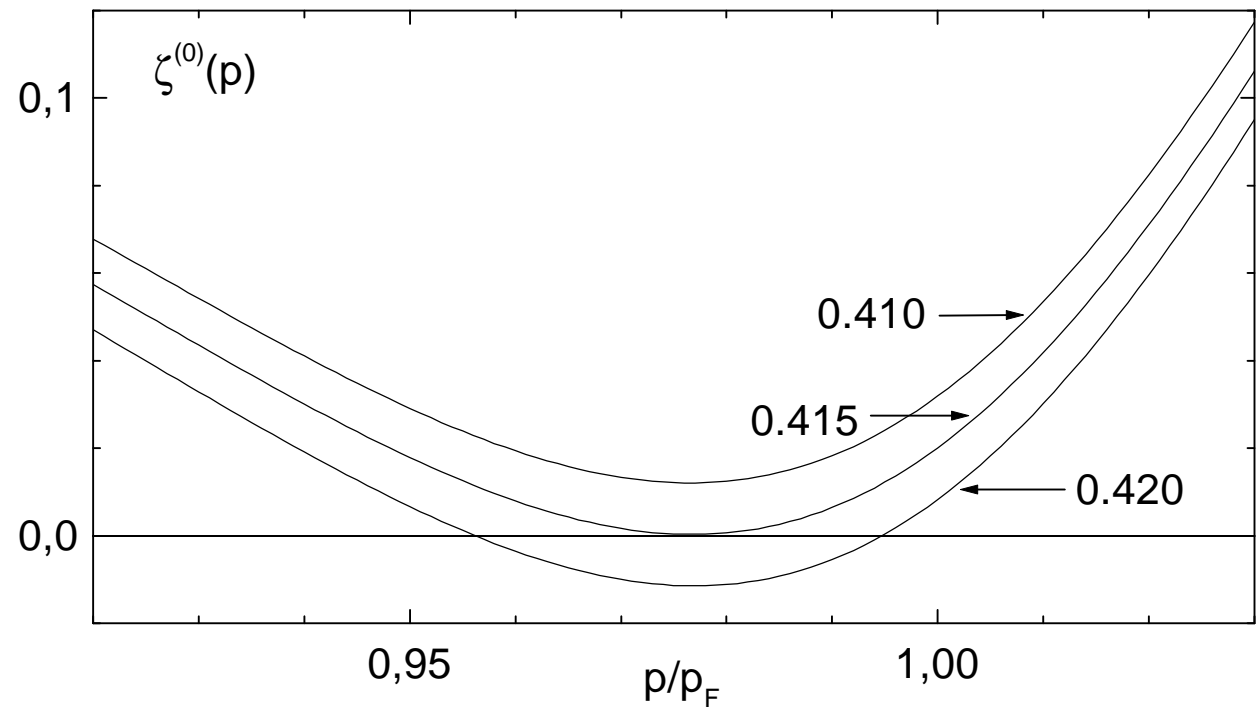
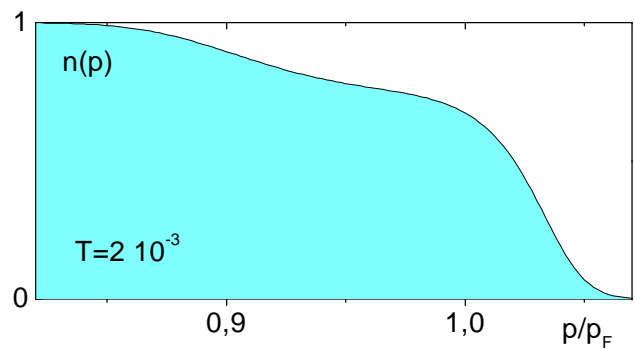
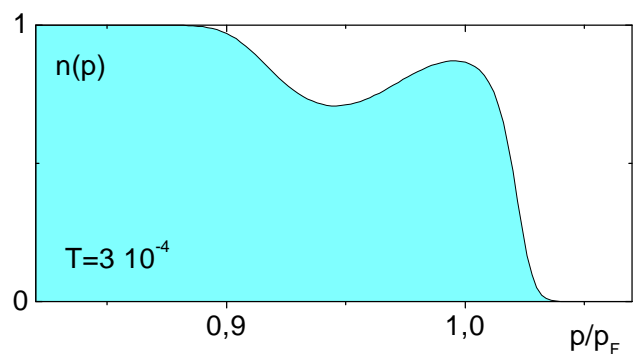
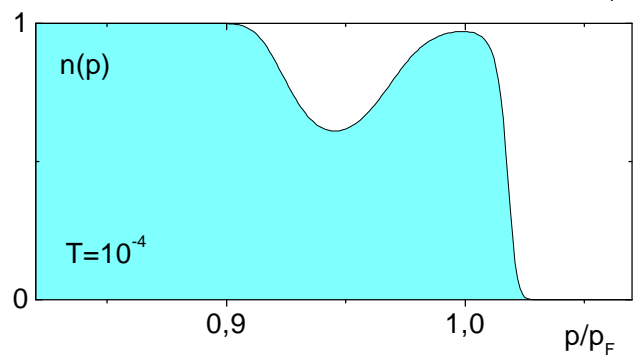
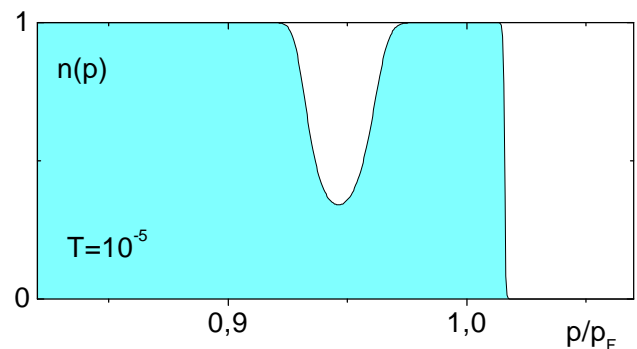
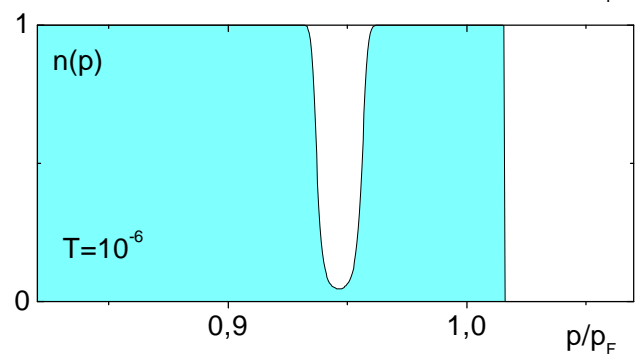
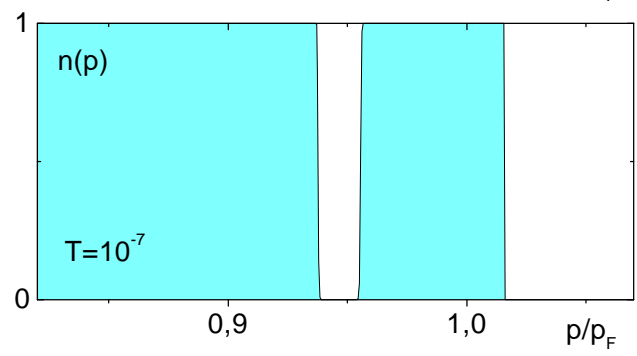
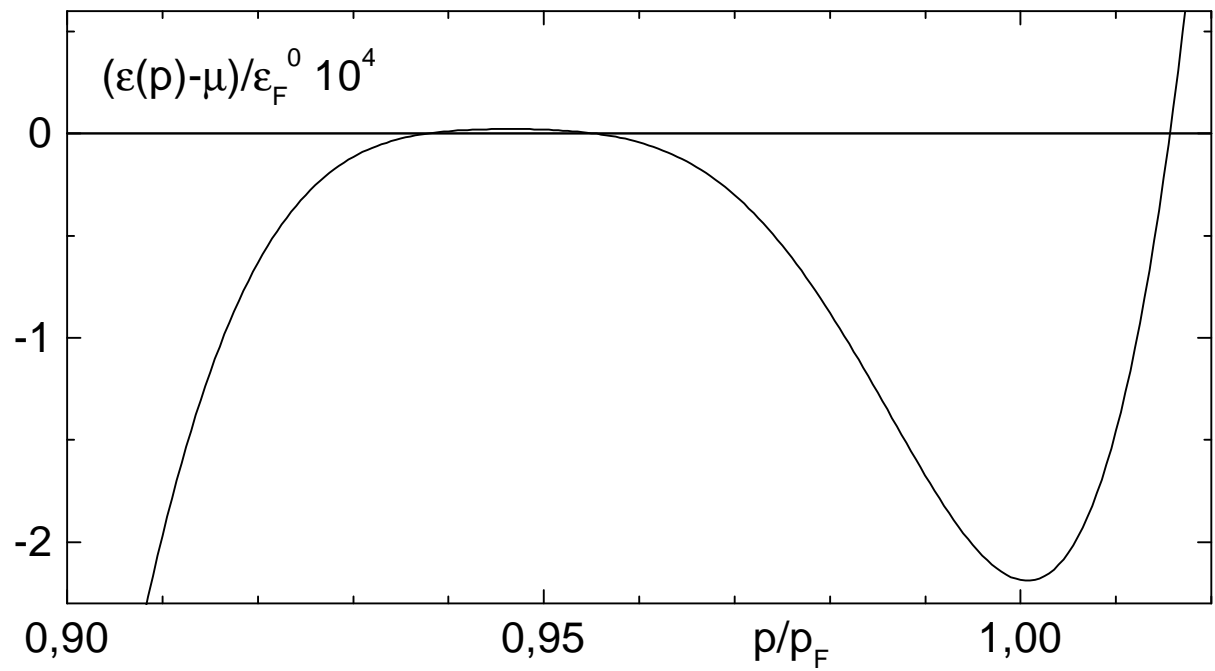
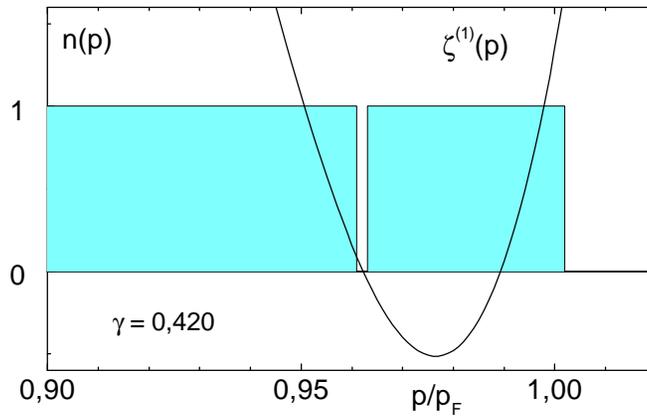
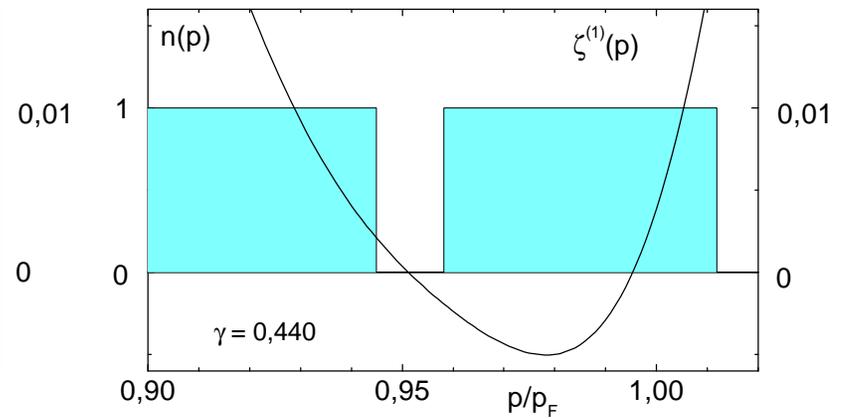
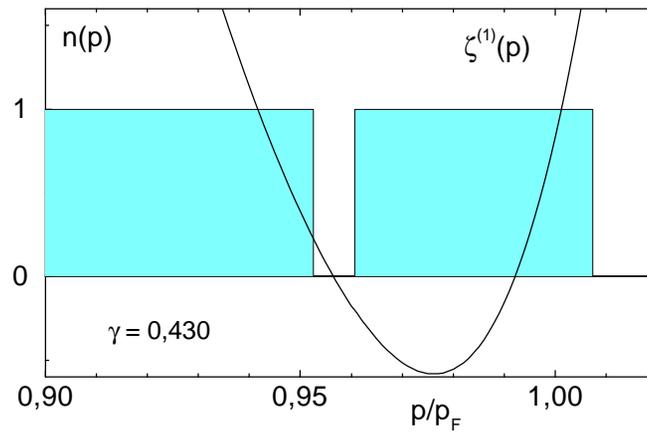
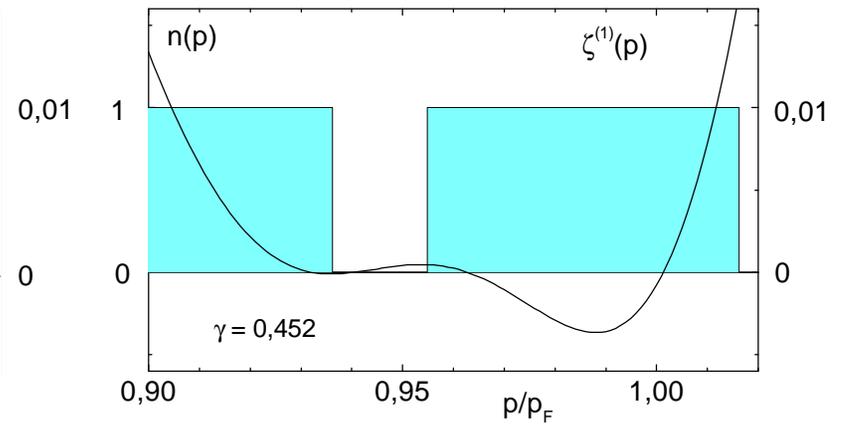
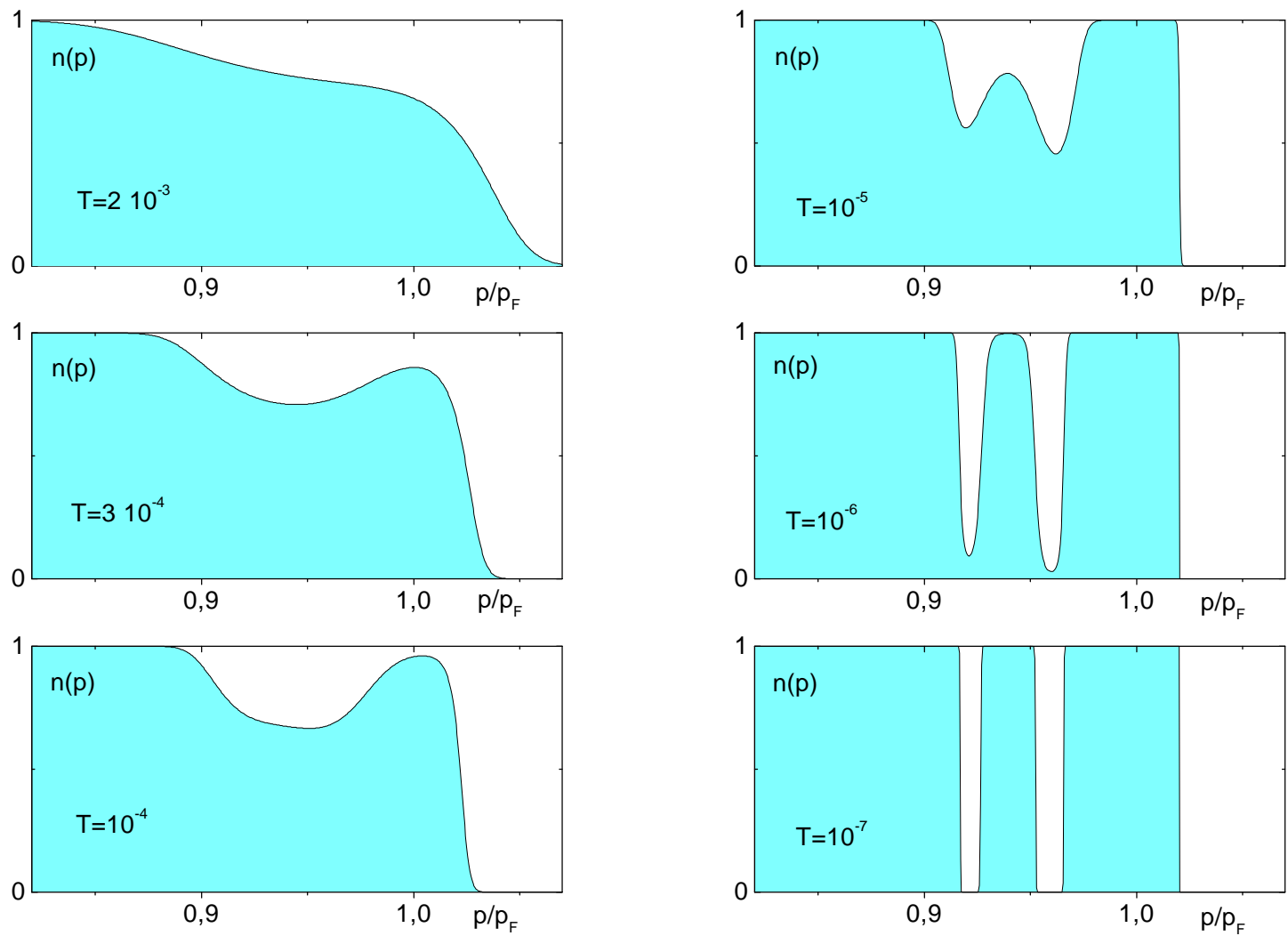
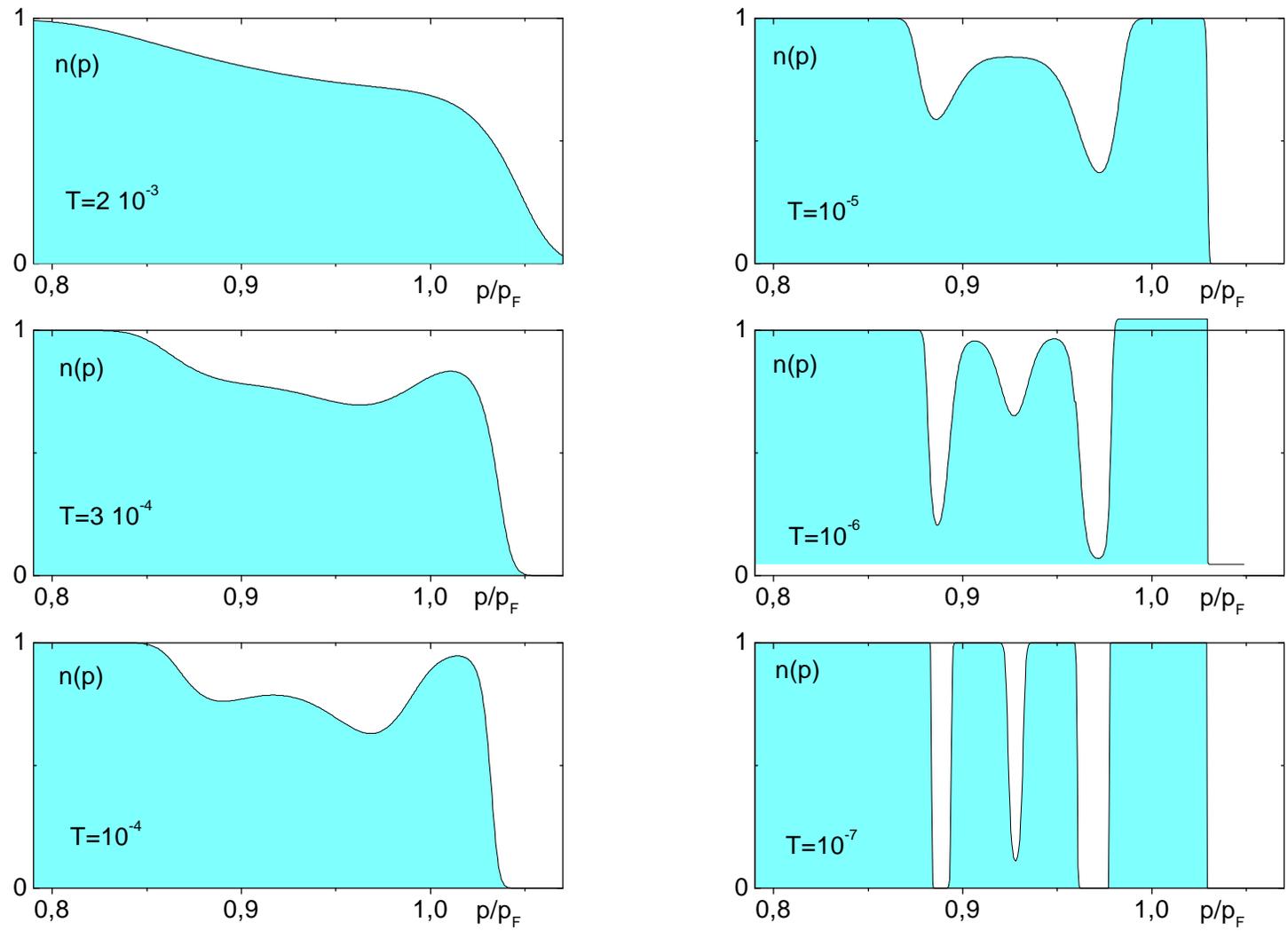
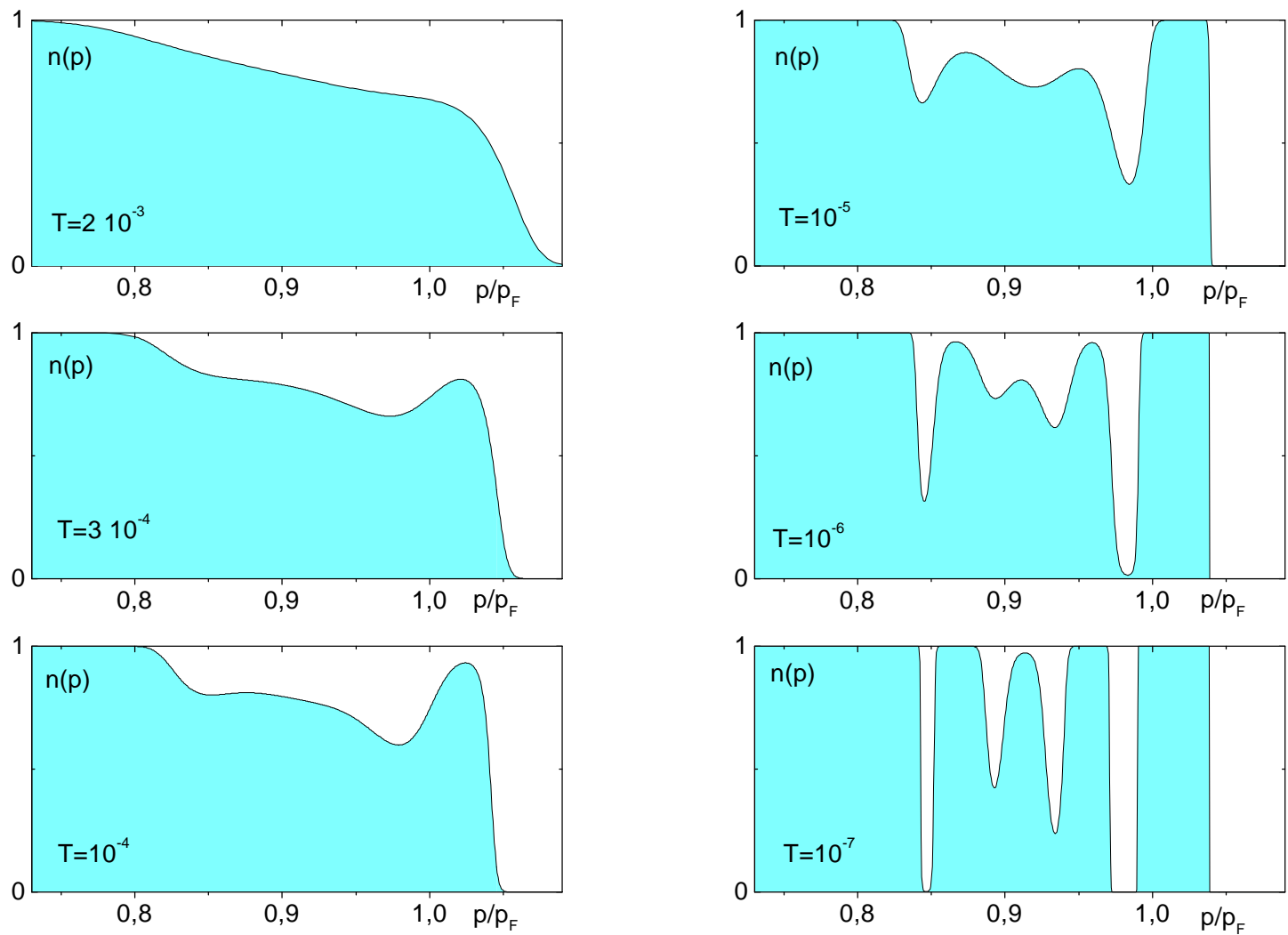
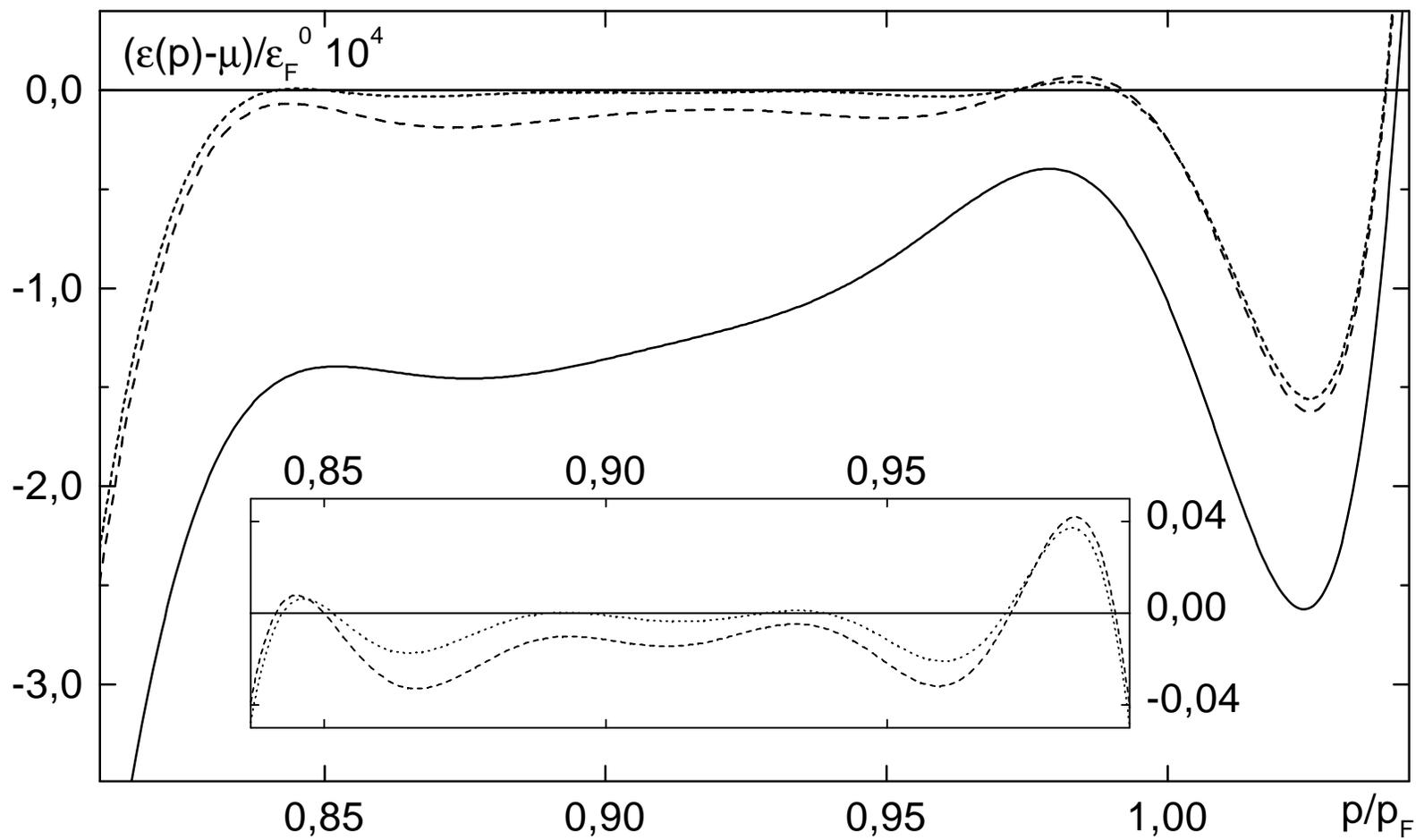
\end{document}